\RequirePackage{fix-cm} 
\documentclass[a4paper, twoside, reqno, 12pt, dvips]{amsart}
\usepackage{fixltx2e}     

\usepackage[english]{babel}
\usepackage[latin1]{inputenc}

\usepackage{eucal}
\usepackage{esint}
\usepackage{amsgen}
\usepackage{amsthm}
\usepackage{amssymb}
\usepackage{amsmath}
\usepackage{upgreek}
\usepackage{amsfonts}
\usepackage{verbatim}
\usepackage[nice]{nicefrac}
\usepackage{mathrsfs, dsfont}

\usepackage{a4wide}
\usepackage{xspace}

\usepackage{indentfirst}
\usepackage{graphicx}
\usepackage{subfigure}
\usepackage[section]{placeins}
\usepackage{psfrag}

\usepackage{microtype}
\usepackage[bf, up, hang]{caption}

\usepackage[usenames, dvipsnames]{pstricks}
\usepackage{epsfig}
\usepackage{pst-grad} 
\usepackage{pst-plot} 

\usepackage{color}
\definecolor{oneblue}{rgb}{0.0, 0.0, 0.85}
\definecolor{darkgrey}{rgb}{0.273, 0.281, 0.30}

\usepackage[colorlinks,
            urlcolor=oneblue,
            linkcolor=oneblue,
            citecolor=oneblue,
            bookmarksopen=false,
            pagebackref]{hyperref}

\usepackage[compact]{titlesec}

\titleformat{\section}{\normalfont\Large\bfseries\sffamily\center\color{darkgrey}}{\thesection.}{0.5em}{}{}
\titleformat{\subsection}{\normalfont\large\bfseries\sffamily\color{darkgrey}}{\thesubsection.}{0.4em}{}{}
\titleformat{\subsubsection}{\normalfont\normalsize\bfseries\sffamily\color{darkgrey}}{\thesubsubsection.}{0.3em}{}{}

\titlespacing*{\section}{1.0em}{1.0em}{0.8em}[0em]
\titlespacing*{\subsection}{1.0em}{1.0em}{0.8em}[0em]
\titlespacing*{\subsubsection}{1.0em}{0.7em}{0.6em}[0em]

\usepackage{titletoc}

\contentsmargin{0.0em}
\dottedcontents{section}[2.5em]{\addvspace{0.45em}\bfseries}{1.0em}{0pc}
\dottedcontents{subsection}[4.5em]{}{2.6em}{0pc}
\dottedcontents{subsubsection}[5.5em]{}{3.0em}{0pc}

\usepackage{fancyhdr}
\usepackage{lastpage}

\newcommand*\Title{Numerical simulation of a solitonic gas}
\newcommand*\Authors{D.~Dutykh \& E.~Pelinovsky}

\usepackage{acronym}

\numberwithin{equation}{section}

\newtheorem{remark}{Remark}

\renewcommand{\k}{\upkappa}

\newcommand{\R}{\mathds{R}}

\renewcommand{\S}{\varsigma}
\newcommand{\St}{\mathrm{S}}
\newcommand{\N}{\mathcal{N}}
\newcommand{\I}{\mathcal{I}}

\newcommand{\eps}{\varepsilon}
\renewcommand{\O}{\mathcal{O}}
\renewcommand{\N}{\mathcal{N}}

\newcommand{\ud}{\mathrm{d}}

\newcommand{\sech}{\mathrm{sech}}

\newcommand{\half}{{\textstyle{1\over2}}}

\newcommand{\third}{{\textstyle{1\over3}}}

\newcommand{\fourth}{{\textstyle{1\over4}}}

\begin{document}

\title[\Title]{Numerical simulation of a solitonic gas in KdV and KdV--BBM equations}

\author[D.~Dutykh]{Denys Dutykh$^*$}
\address{LAMA, UMR 5127 CNRS, Universit\'e de Savoie, Campus Scientifique, 73376 Le Bourget-du-Lac Cedex, France}
\email{Denys.Dutykh@univ-savoie.fr}
\urladdr{http://www.denys-dutykh.com/}
\thanks{$^*$ Corresponding author}

\author[E.~Pelinovsky]{Efim Pelinovsky}
\address{Department of Nonlinear Geophysical Processes, Institute of Applied Physics, Nizhny Novgorod, Russia \and Department of Applied Mathematics, Nizhny Novgorod State Technical University, Russia \and National Research University --- Higher School of Economics, Russia}
\email{pelinovsky@hydro.appl.sci-nnov.ru}
\urladdr{http://www.ipfran.ru/english/staff/Pelinovsky.html}

\begin{abstract}
The collective behaviour of soliton ensembles (i.e. the solitonic gas) is studied using the methods of the direct numerical simulation. Traditionally this problem was addressed in the context of integrable models such as the celebrated KdV equation. We extend this analysis to non-integrable KdV--BBM type models. Some high resolution numerical results are presented in both integrable and nonintegrable cases. Moreover, the free surface elevation probability distribution is shown to be quasi-stationary. Finally, we employ the asymptotic methods along with the Monte--Carlo simulations in order to study quantitatively the dependence of some important statistical characteristics (such as the kurtosis and skewness) on the Stokes--Ursell number (which measures the relative importance of nonlinear effects compared to the dispersion) and also on the magnitude of the BBM term.

\bigskip
\noindent \textbf{\keywordsname:} solitonic gas; KdV equation; BBM equation; statistical description; statistical moments; Stokes--Ursell number
\end{abstract}

\subjclass[2010]{76B25, 76B15, 35Q53}

\maketitle
\tableofcontents
\thispagestyle{empty}

\section{Introduction}

Solitary wave solutions play the central r\^ole in various nonlinear sciences ranging from hydrodynamics to solid and plasma physics \cite{Zabusky1965, Salupere2003, Osborne2010}. These solutions can propagate without changing its shape. However, the most intriguing part consists in how these solutions interact with each other. The binary interactions of solitary waves have been studied in the context of various nonlinear wave equations \cite{Zabusky1965, Maxworthy1976, Su1980, Mirie1982, CGHHS, Chambarel2009}. It is well known that in integrable models the collision of two solitons is elastic, \emph{i.e.} they interact without emitting any radiation. In non-integrable models usually the interactions are nearly elastic \cite{Bogolubsky1977}.

The collective behaviour of soliton ensembles is much less understood nowadays. When a large number of solitary waves are considered simultaneously the researchers usually speak about the so-called solitonic turbulence or a solitonic gas. The literature on this topic is abundant. Some recent studies on solitonic gas turbulence in the KdV framework include \cite{Pelinovsky2006, Sergeeva2011, Pelinovsky2013}. The solitonic turbulence in nonintegrable NLS-type equations was studied in \cite{Zakharov1988, Dyachenko1989} and the authors showed that in conservative nonintegrable systems the solitonic gas is a statistical attractor whose dimension decreases with time. Recently it was shown both numerically and experimentally that solitonic ensembles appear in the laminar--turbulent transition in a fibre laser \cite{Turitsyna2013}, modeled by a non-integrable nonlinear Schr\"odinger-type equation. However, the dominant number of studies is based on integrable models. This apparent contradiction motivated mainly our investigation to quantify the non-integrable effects onto the collective behaviour of solitons.

An approximate theoretical description of solitonic gases was proposed by V.~\textsc{Zakharov} (1971) \cite{Zakharov1971} using the kinetic theory. Later this research direction has been successfully pursued by G.~\textsc{El} \& A.~\textsc{Kamchatnov} \cite{El2005a, El2011} who used the Inverse Scattering Technique (IST) \cite{Ablowitz1981} limited only to the integrable models. In this study the problem of solitonic gases will be investigated using the methods of direct numerical simulation. The evolution of random wave fields including solitonic gases was simulated numerically in \cite{Pelinovsky2006, Sergeeva2011, Dutykh2013a} using symplectic, multi-symplectic and pseudo-spectral methods. However, previous investigators considered only a limited number of solitons (a few dozens) to simulate a solitonic gas. In this study we will adopt the pseudo-spectral method since it provides the high accuracy and computational efficiency necessary to handle large computational domains. Our goal will consist in:
\begin{itemize}
  \item investigate the influence of soliton interactions on statistical characteristics of the wave field,
  \item construct the Probability Density Function (PDF) and compute the first four statistical moments of the solitonic turbulence,
  \item study the r\^ole of non-integrable terms on the characteristics of soliton ensembles.
\end{itemize}

The present manuscript is organized as follows. In Section~\ref{sec:model} we derive the governing equation used in this study and in Section~\ref{sec:num} numerical results on a solitonic gas dynamics are presented. Finally, the main conclusions of this study are outlined in Section~\ref{sec:concl}.

\section{Mathematical model}\label{sec:model}

As the starting point we choose the celebrated Korteweg--de Vries equation \cite{KdV, Miles1981, Johnson2002, Pelinovsky2006} (in dimensional variables) which models the undirectional propagation (here in the rightwards direction) of weakly nonlinear and weakly dispersive waves:
\begin{equation}\label{eq:kdv}
  \eta_t + c\Bigl(1 + \frac{3h}{2}\eta\Bigr)\eta_x + \frac{ch^2}{6}\eta_{xxx} = 0,
\end{equation}
where $\eta(x,t)$ is the vertical excursion of the free surface above the still water level, $h$ is the uniform undisturbed water depth and $c = \sqrt{gh}$ is the speed of linear gravity waves ($g$ being the gravity acceleration).

The KdV equation \eqref{eq:kdv} is known to be integrable \cite{Gardner1967, Miura1968}. However, the full water wave problem is known to be a non-integrable system, since the interaction of solitary waves is inelastic \cite{Chan1970, Fenton1982, Cooker1997, CGHHS}. Moreover, we will modify the original equation \eqref{eq:kdv} in order to include an additional dispersive term of the BBM-type \cite{bona}. Thus, the resulting KdV--BBM equation will not be integrable \cite{Francius2001, Dutykh2010e, Kalisch2013}.

Consider the following scaled dependent and independent variables:
\begin{equation*}
  \eta \gets \frac{\eta}{a_0}, \quad x \gets \frac{x}{l}, \quad
  t \gets \frac{ct}{l},
\end{equation*}
where $a_0$ is the characteristic wave amplitude and $l$ is the characteristic wavelength. In dimensionless variables KdV equation \eqref{eq:kdv} reads:
\begin{equation*}
  \eta_t + \Bigl(1 + \frac{3\eps}{2}\eta\Bigr)\eta_x + \frac{\mu^2}{6}\eta_{xxx} = 0,
\end{equation*}
where parameter $\eps := \displaystyle{\nicefrac{a_0}{h}}$ measures the nonlinearity and $\mu^2 := \bigl(\displaystyle{\nicefrac{h}{l}}\bigr)^2$ is the dispersion parameter. The relative importance of these two effects is measured by the so-called Stokes--Ursell number \cite{Ursell1953} (sometime denoted as $\mathrm{Ur}$):
\begin{equation*}
  \St := \frac{\eps}{\mu^2} \equiv \frac{a_0 l^2}{h^3}.
\end{equation*}
The last equation can be further simplified if we perform an additional change of variables:
\begin{equation*}
  \eta \gets \frac{3\mu^2}{2}\eta, \quad x \gets \frac{\sqrt{6}}{\mu}(x-t), \quad t \gets \frac{\sqrt{6}}{\mu}t,
\end{equation*}
which yields the following simple equation including explicitly the Stokes--Ursell number $\St$:
\begin{equation}\label{eq:kdv}
  \eta_t + \St\eta\eta_x + \eta_{xxx} = 0.
\end{equation}
The last scaled KdV equation can be further generalized by using the low-order asymptotic relations in order to alternate higher-order terms as it was proposed by \textsc{Bona} \& \textsc{Smith} (1976) \cite{BS} and \textsc{Nwogu} (1993) \cite{Nwogu1993}. This step is rather standard and we do not provide here the details of the derivation:
\begin{equation}\label{eq:kdvbbm}
  \eta_t + \St\eta\eta_x + \eta_{xxx} - \delta\eta_{xxt} = 0,
\end{equation}
where $\delta\in\R$ is a free parameter. The solitary wave collisions in this equation were studied earlier by \textsc{Francius} \emph{et al.} (2001) \cite{Francius2001} and \textsc{Kalisch} \emph{et al.} (2013) \cite{Kalisch2013}. Below we will study the solitonic gas behaviour\footnote{Sometimes it is also called the solitonic turbulence, e.g. in \cite{Zakharov1988, Dyachenko1989}.} under the dynamics of the KdV \eqref{eq:kdv} ($\delta = 0$) and  KdV--BBM \eqref{eq:kdvbbm} ($\delta \neq 0$) equations. In the absence of the KdV term, we recover the celebrated BBM equation derived in \cite{Peregrine1966, bona}.

\begin{remark}
We note that for a particular value of the Stokes--Ursell number $\St \equiv 1$ another simpler scaling is possible when all the lengthes ($x$ and $\eta$) are scaled by the mean water depth $h$. However, we do not adopt it in this study since below in Section~\ref{sec:stat} the dependence of some statistical characteristics on the Stokes--Ursell number $\St$ is investigated.
\end{remark}

\subsection{Properties}

\subsubsection{Linear well-posedness}

In order to ensure the linear well-posedness of equation \eqref{eq:kdvbbm}, the free parameter $\delta$ has to satisfy the following constraint $\delta\geq 0$. In the following we will consider only nonnegative values of this parameter. Recall that for $\delta = 0$ we recover an scaled version of the classical KdV equation \eqref{eq:kdv}.

\subsubsection{Ivariants}

Assuming that the solution $\eta(x,t)$ has either the compact support or decays sufficiently fast at the infinity along with its first derivative ($\eta\to 0$, $\eta_x\to 0$ as $x\to\pm\infty$), one can easily show that the following quantities are conserved \cite{Dutykh2010e}:
\begin{equation*}
  \I_1(t) = \int_\R\eta(x,t)\,\ud x, \qquad
  \I_2(t) = \int_\R\Bigl(\eta^2(x,t) + \delta\eta_x^2(x,t)\Bigr)\,\ud x.
\end{equation*}
In other words, $\I_1(t) \equiv \I_1(0)$ and $\I_2(t) \equiv \I_2(0)$, $\forall t > 0$. The invariant $\I_1(t)$ is related to the mass conservation property, while the integral $\I_2(t)$ can be assimilated to the generalized kinetic energy. The conservation of these quantities has not only the theoretical importance, but also the practical one. For example, it will allow us to check the global accuracy of the employed numerical scheme. We note also that the same invariants hold also in finite, but periodic domains (below we use periodic boundary conditions). The conservation laws to the BBM equation can be found in \cite{Olver1979}.

\subsubsection{Solitary wave solutions}

\begin{figure}
  \centering
  \subfigure[]{%
  \includegraphics[width=0.49\textwidth]{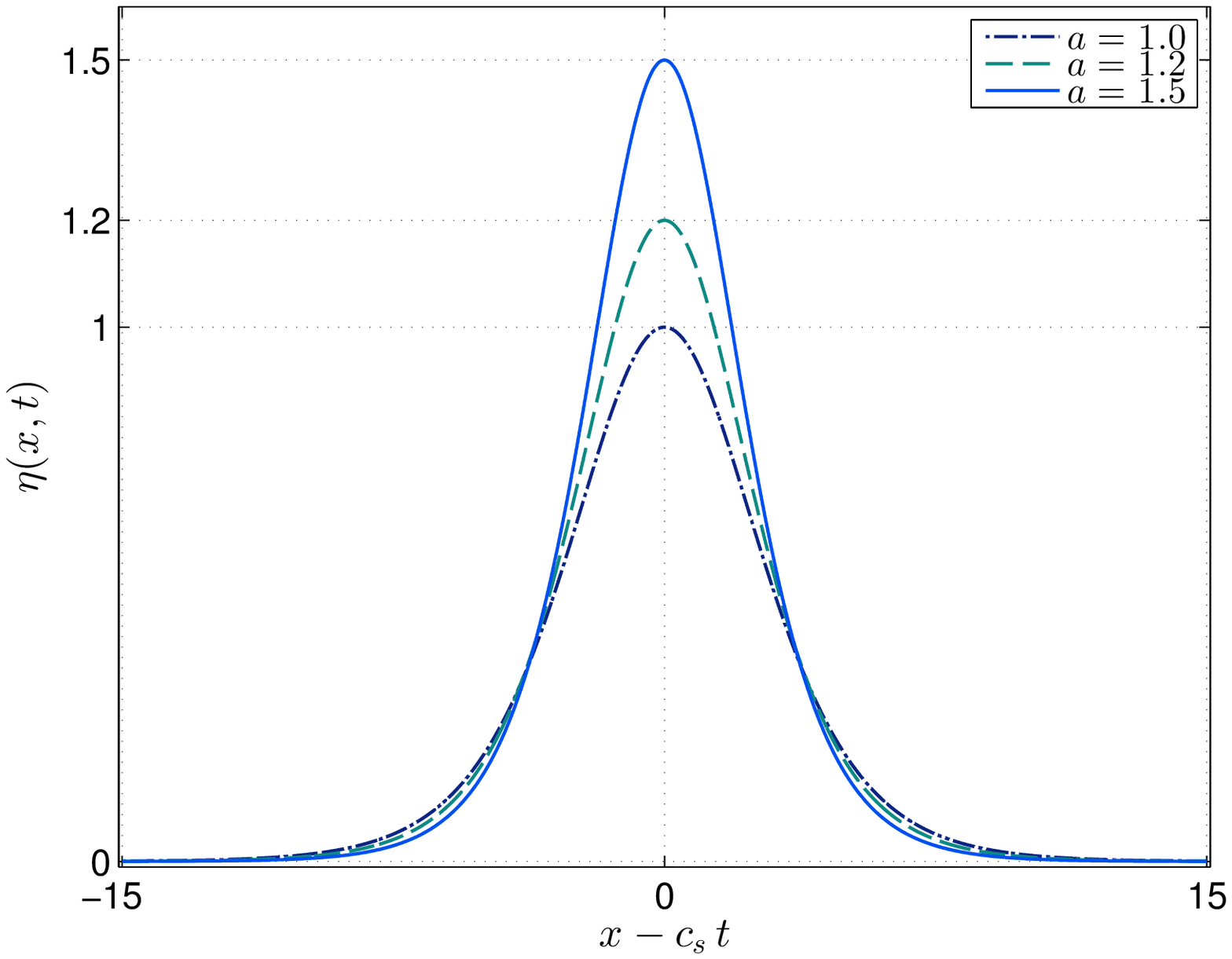}}
  \subfigure[]{%
  \includegraphics[width=0.49\textwidth]{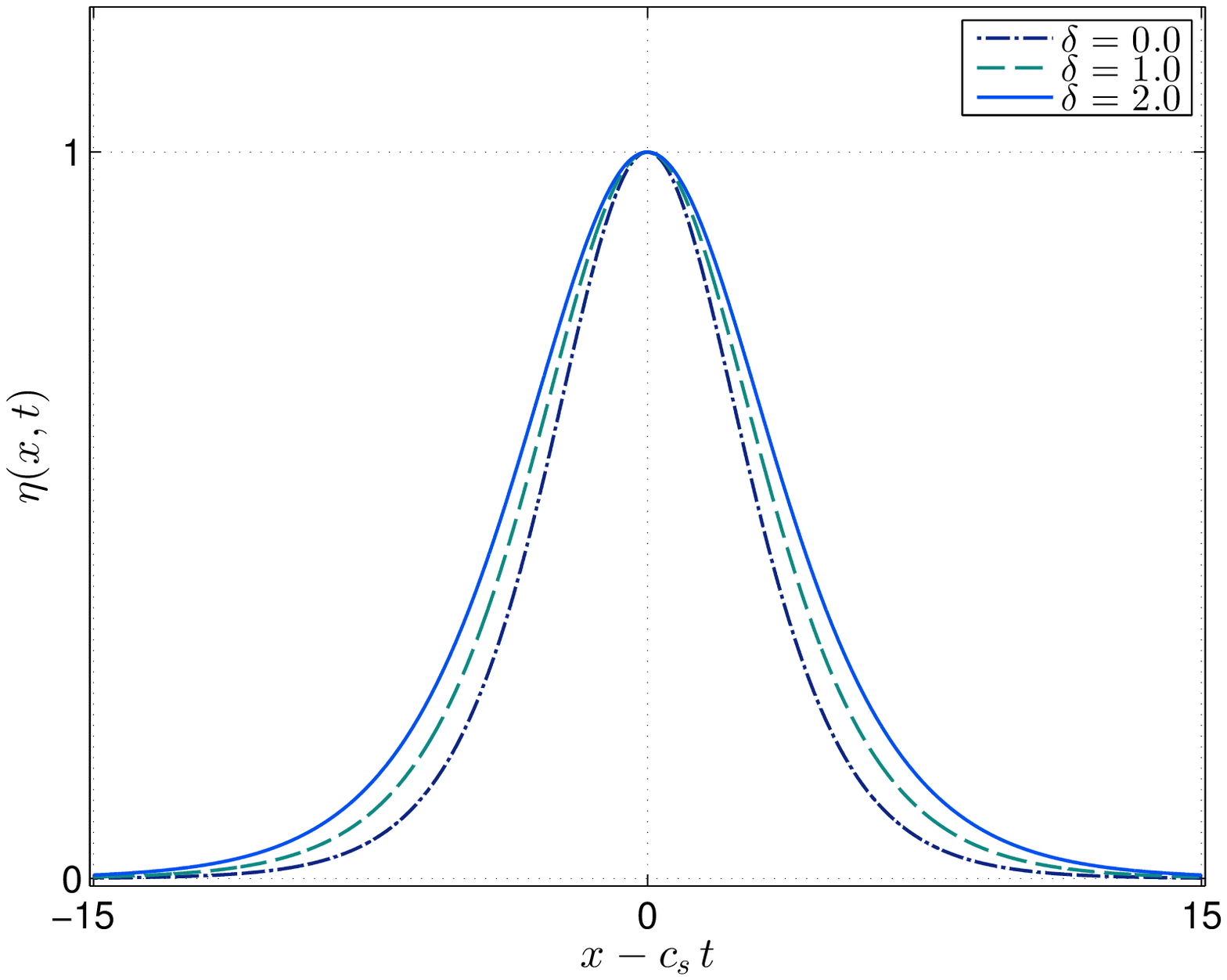}}
  \subfigure[]{%
  \includegraphics[width=0.49\textwidth]{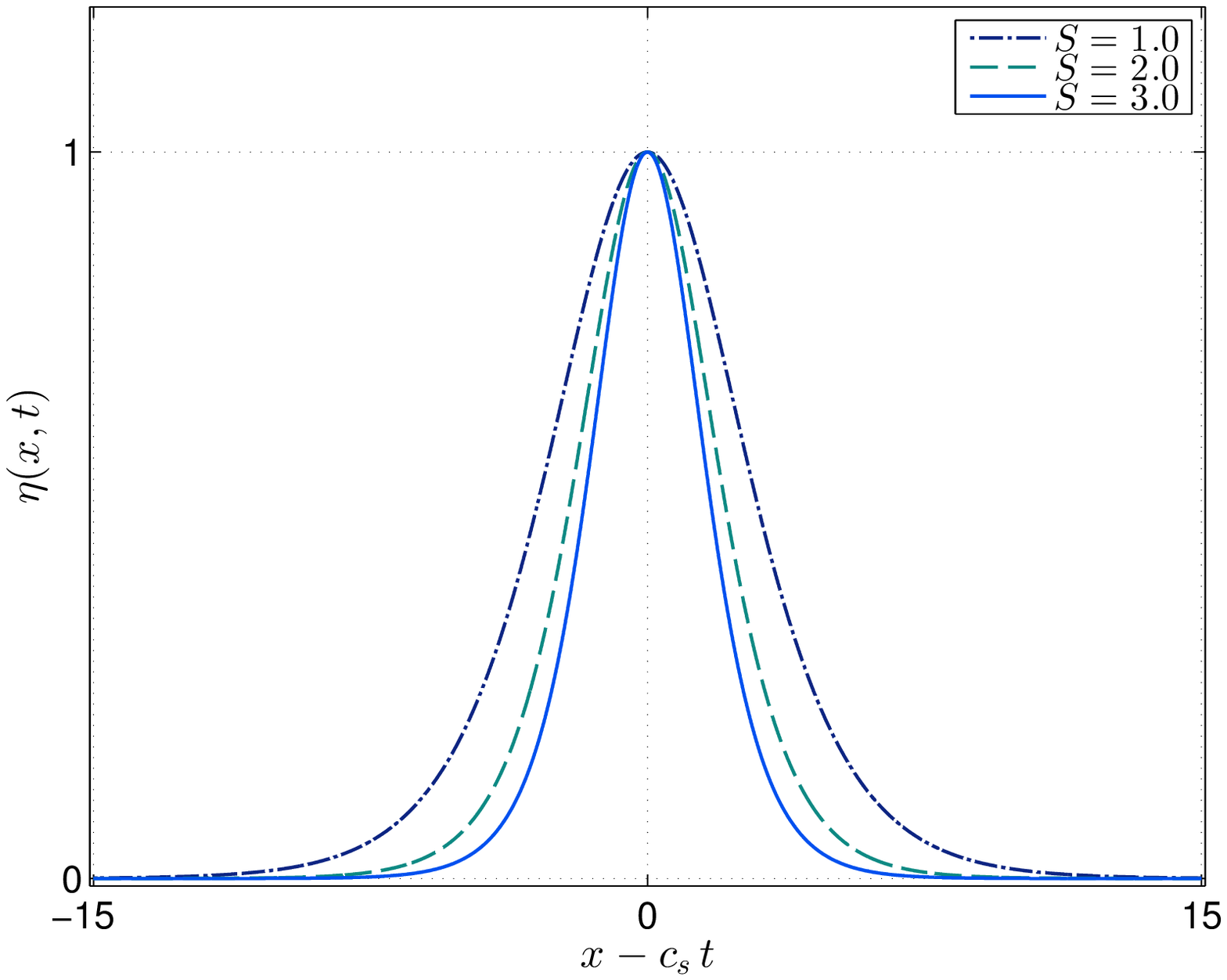}}
  \subfigure[]{%
  \includegraphics[width=0.49\textwidth]{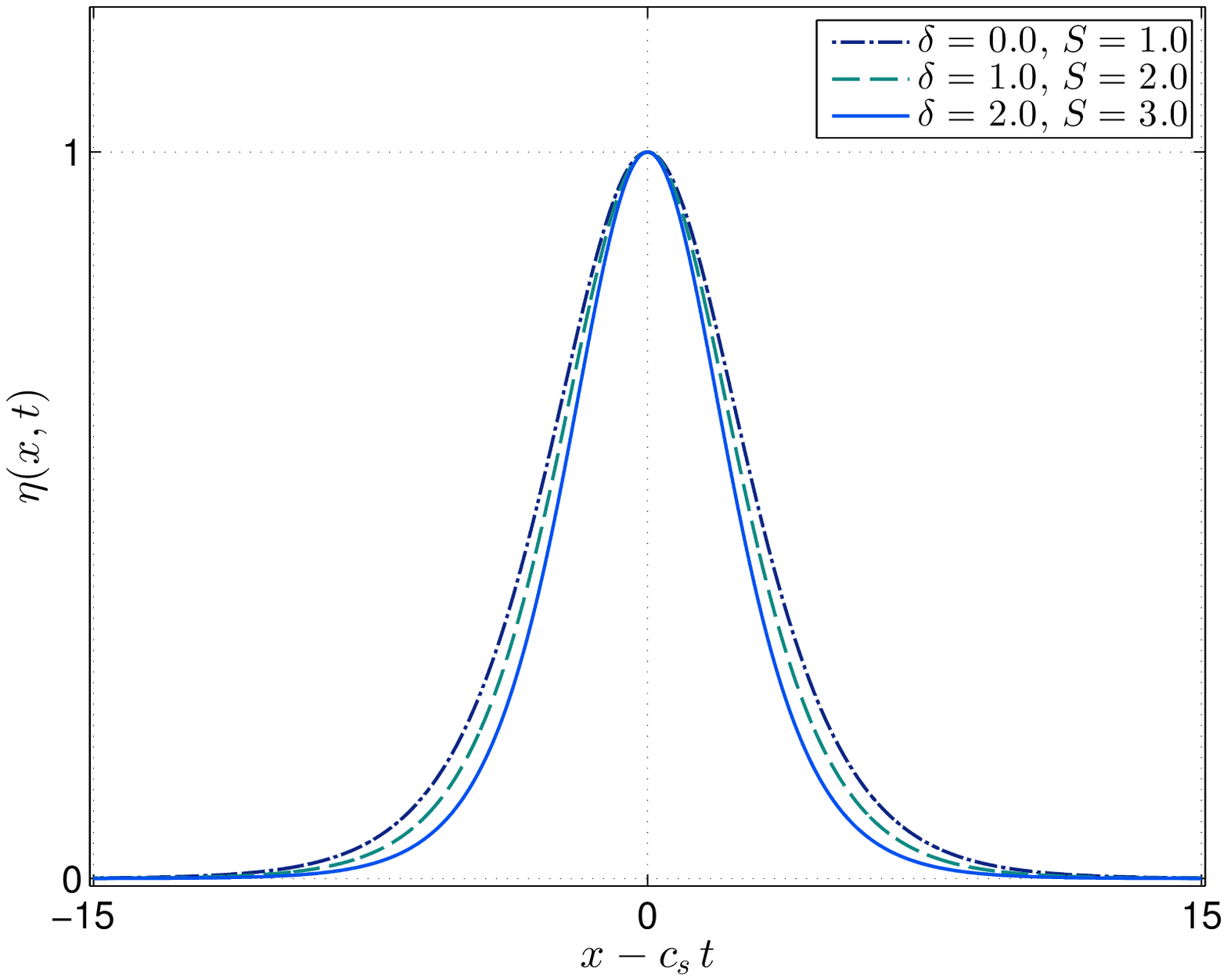}}
  \caption{\small\em Solitary wave shape dependence on various parameters: (a) amplitude $a = 1.0$, $1.2$, $1.5$ for $\St=1.0$, $\delta = 0$ (b) $\delta = 0.0$, $1.0$, $2.0$ for $a = 1.0$, $\St = 1.0$ (c) $\St = 1.0$, $2.0$, $3.0$ for $a = 1.0$, $\St = 1.0$ (d) simultaneous change of $(\delta, \St) = (0.0, 1.0)$, $(1.0, 2.0)$, $(2.0, 3.0)$ for $a = 1.0$.}
  \label{fig:shapes}
\end{figure}

Equation \eqref{eq:kdvbbm} admits an exact localized (solitary) travelling wave solution which can be found analytically:
\begin{equation}\label{eq:sw}
  \eta(x,t) = a\,\sech^2\bigl(\half\kappa(x-c_st)\bigr), \qquad \kappa := \sqrt{\frac{a\St}{3 + a\St\delta}}, \qquad c_s := \third a\St.
\end{equation}
The dependence of the solitary wave shape on parameters $a$, $\St$ and $\delta$ is shown on Figure~\ref{fig:shapes}. For instance, one can see that solitary waves become thiner when the amplitude $a$ (and hence the speed) and/or the Stokes--Ursell number $\St$ are increased (see Figure~\ref{fig:shapes}(a,c)). On the other hand, the increase of the BBM coefficient $\delta$ leads to the growth of the tail (see Figure~\ref{fig:shapes}(b)). If both parameters $\St$ and $\delta$ are increased simultaneously, the `thinning' effect of the Stokes--Ursell number dominates (see Figure~\ref{fig:shapes}(d)).

The solitary waves interact elastically only in the integrable KdV case ($\delta \equiv 0$) \cite{Miura1976}, while some dispersive tails are generated after the interaction in the general KdV--BBM model. We note that some authors came to the wrong conclusion about the elasticity of solitary wave interactions in the BBM equation \cite{Eilbeck1977} on the basis of low accuracy numerical simulations.

\section{Numerical results}\label{sec:num}

In order to solve numerically equation \eqref{eq:kdvbbm} we use a Fourier-type pseudo-spectral method with $3/2$-antialiasing rule \cite{Trefethen2000}. For the time discretization we use the Verner's embedded adaptive 9(8) Runge--Kutta scheme \cite{Verner1978}. The time step is chosen adaptively using the so-called \textsc{H211b} digital filter \cite{Soderlind2003, Soderlind2006} to meet some prescribed error tolerance (generally of the order of machine precision $\sim 10^{-15}$). The number of Fourier modes, the length of the computational domain and other numerical parameters are specified in Table~\ref{tab:params}.

\begin{table}
  \centering
  \begingroup\setlength{\fboxsep}{0pt}
  \colorbox{lightgray}{
  \begin{tabular}{l|c}
  \hline\hline
  Stokes--Ursell number: $\St$ & $1.0$ \\
  BBM term coefficient: $\delta$ & $0.0$ ($2.0$) \\
  Number of Fourier modes: $N$ & $2^{17} = 131072$ \\
  Half-length of the domain $[-\ell, \ell]$: $\ell$ & $4558.0$ ($5580.0$) \\
  Final simulation time: $T$ & $30 000$ ($35 000$) \\
  Average time step: $\Delta t$ & $0.02$ ($0.0194$) \\
  Number of solitons in the gas: $N_s$ & 200 \\
  Average distance between solitons: $\langle\Delta x_s\rangle$ & $45.0$ ($55.0$) \\
  Average amplitude of a soliton: $a$ & $1.0$ \\
  Variance of soliton amplitude: $\sigma$ & $0.2$ \\
  Variance of the soliton position: $\sigma_2$ & $4.0$ \\
  Number of Monte--Carlo realizations: $M$ & 100 \\
  \hline\hline
  \end{tabular}}\endgroup
  \bigskip
  \caption{\small\em Physical and numerical parameters used for simulations of the solitonic gas in the KdV and KdV--BBM (in parentheses) dynamics.}
  \label{tab:params}
\end{table}

\begin{figure}
  \centering
  \subfigure[]{%
  \includegraphics[width=0.49\textwidth]{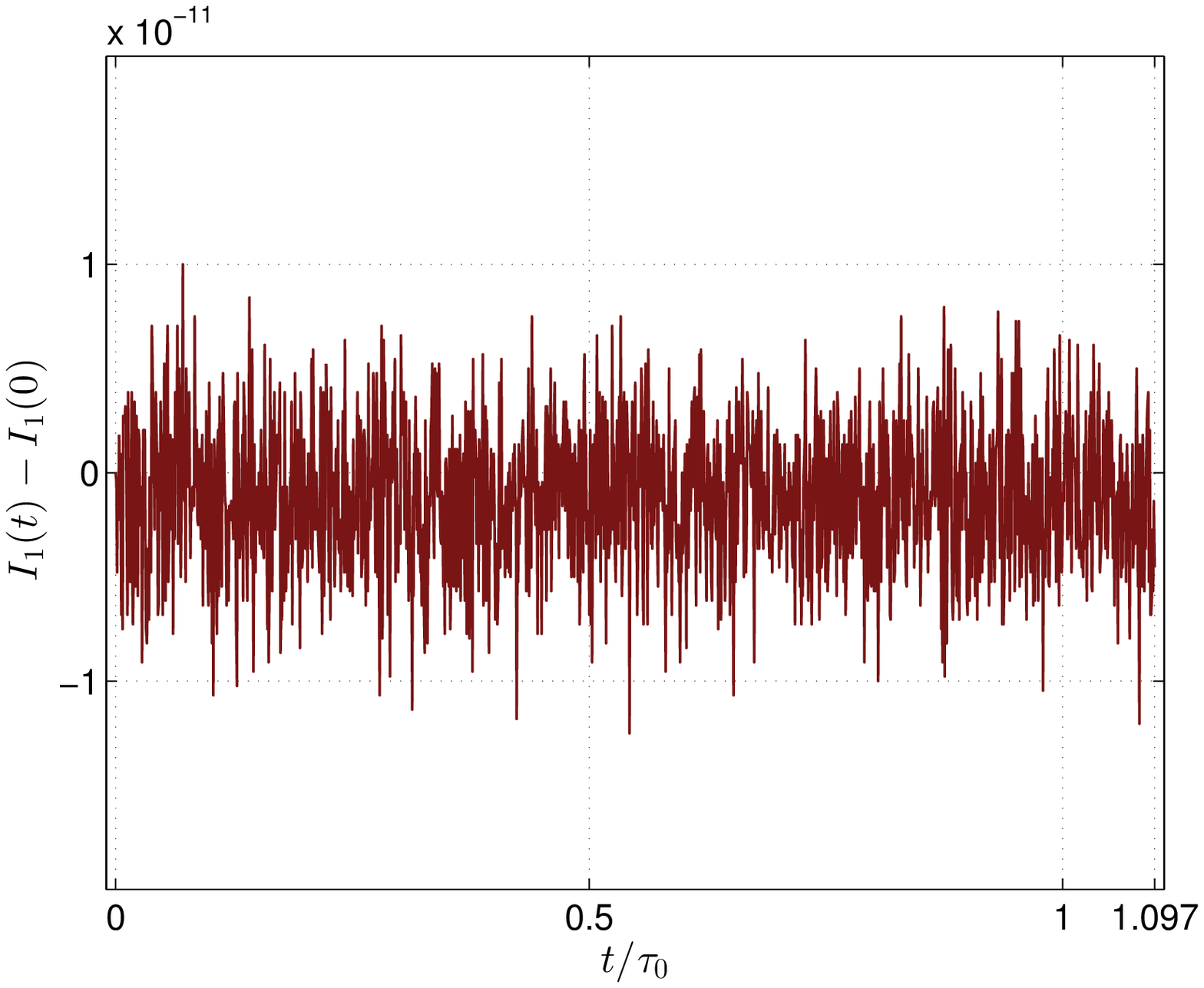}}
  \subfigure[]{%
  \includegraphics[width=0.49\textwidth]{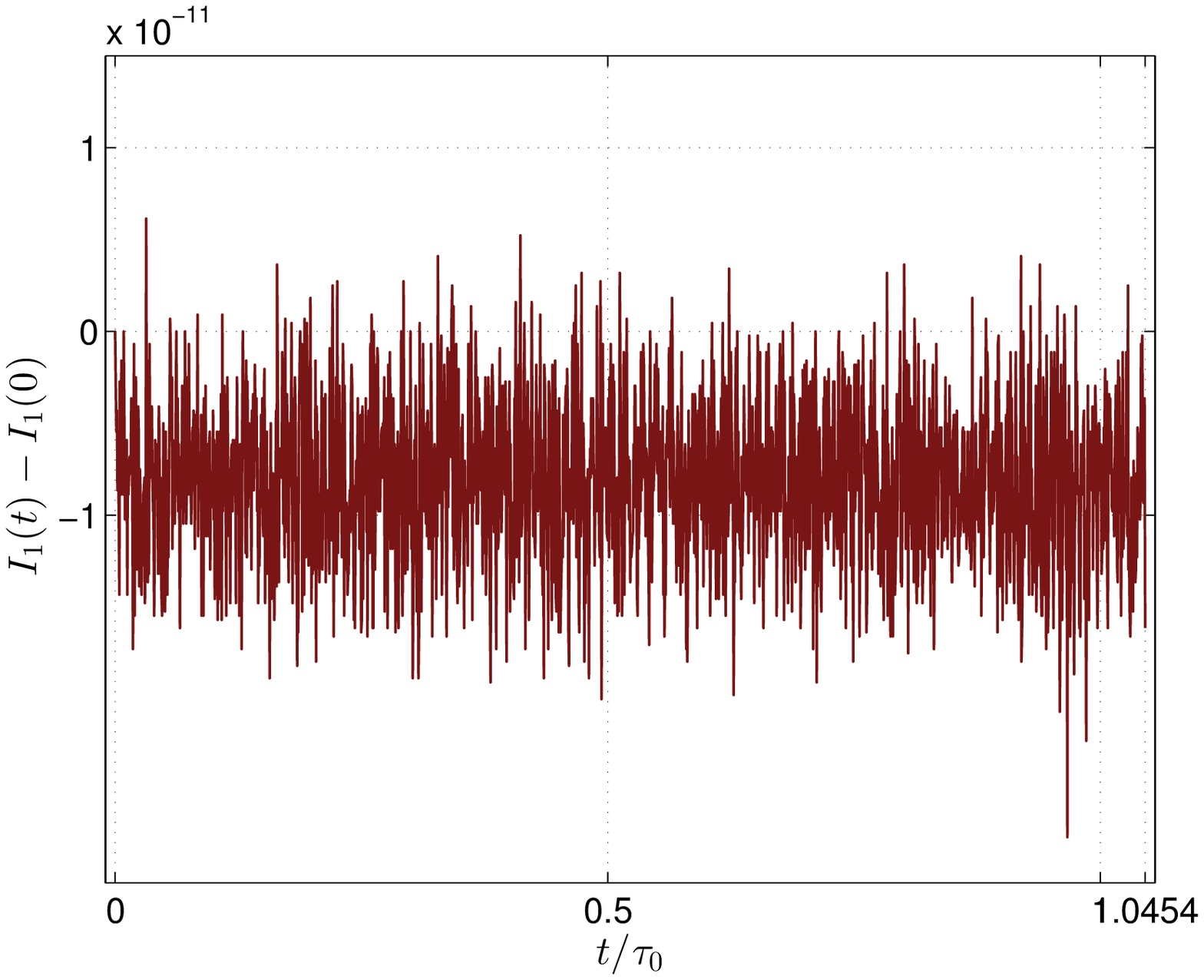}}
  \caption{\small\em Conservation of the invariant $\I_1(t)$ during the KdV (a) and KdV--BBM (b) simulations.}
  \label{fig:i1}
\end{figure}

In long time simulations presented below the invariants $\I_{1,2}$ were conserved within the numerical accuracy $10^{-11}$ and $10^{-9}$ correspondingly. For the sake of illustration the error of the invariant $\I_1(t)$ conservation in KdV and KdV--BBM simulations is shown on Figure~\ref{fig:i1}. This accuracy is satisfactory to draw some robust conclusions on a solitonic gas statistical behaviour.

\begin{figure}
  \centering
  \subfigure[$t=0$, KdV]{%
  \includegraphics[width=0.99\textwidth]{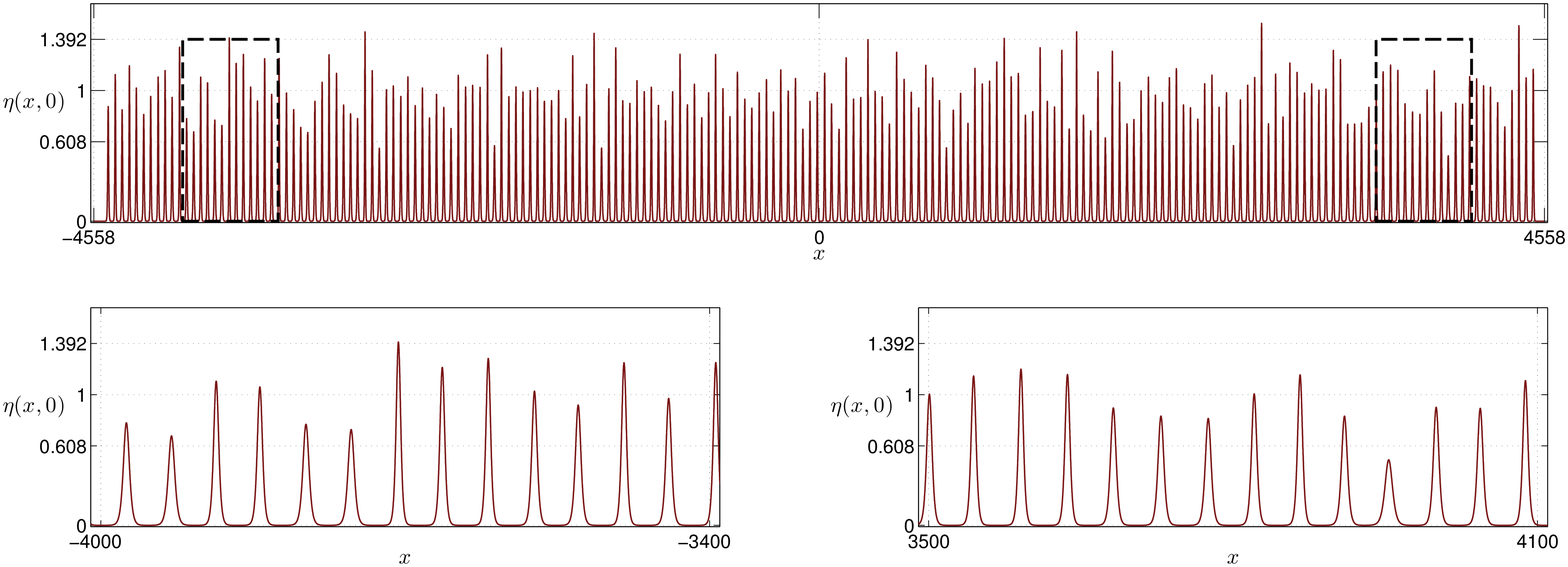}}
  \subfigure[$t = T_1$, KdV]{%
  \includegraphics[width=0.99\textwidth]{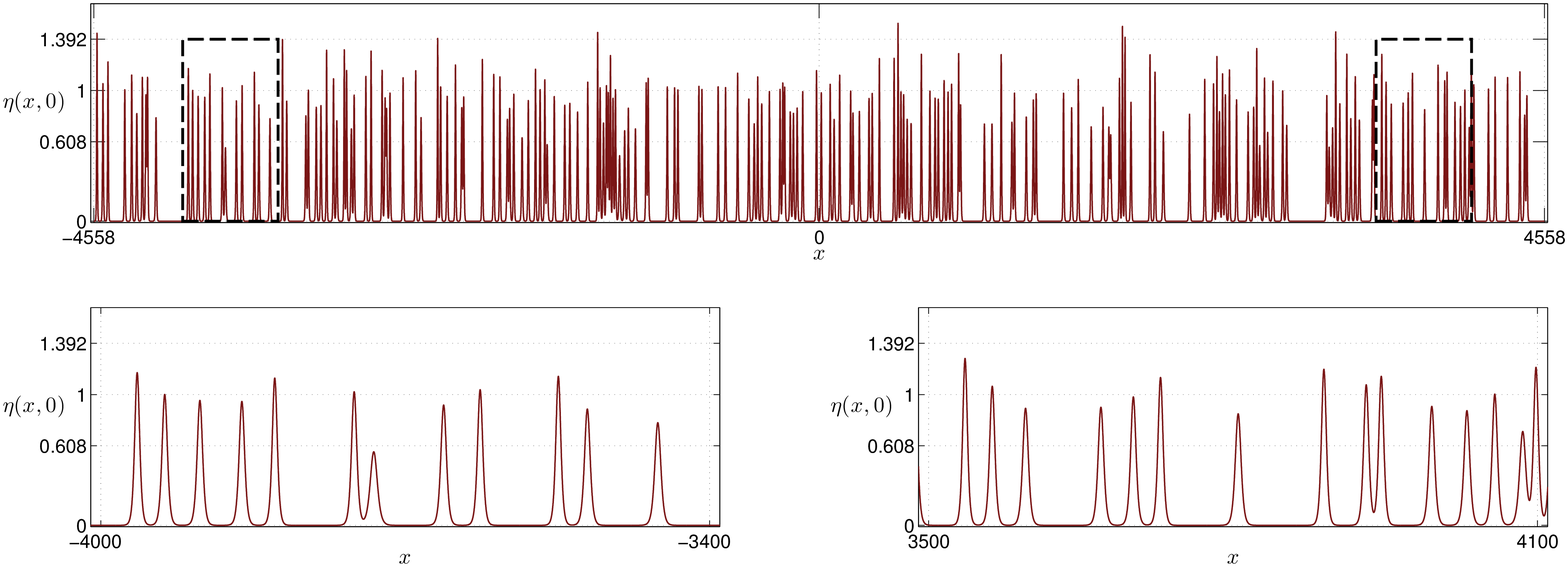}}
  \subfigure[$t = T_2$, KdV--BBM]{%
  \includegraphics[width=0.99\textwidth]{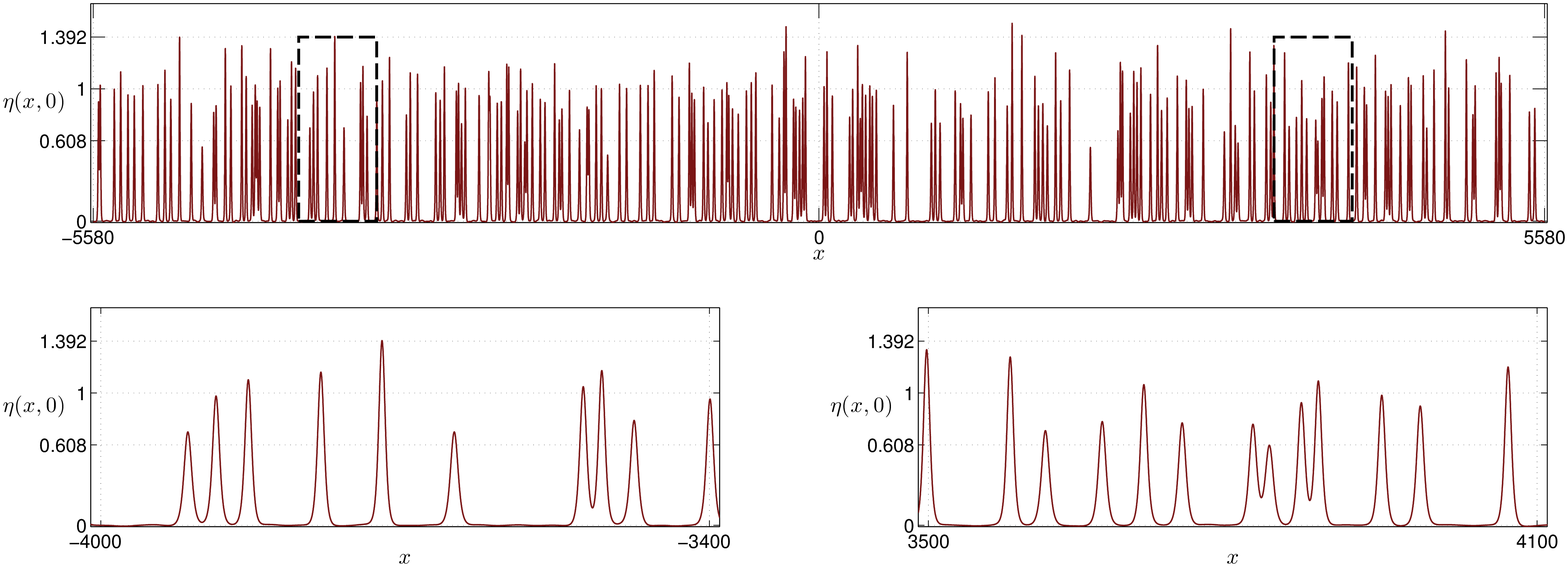}}
  \caption{\small\em The initial condition (a) and the final state (b) of a random solitonic gas simulated using the KdV equation ($\delta = 0$). The final state of the KdV--BBM simulation is shown on panel (c). Please, note that the final simulation times $T_1 \neq T_2$ (see Table~\ref{tab:params} for the values of $T_{1,2}$). Parameter $\tau_0$ denotes the time needed for an average soltion to go over the whole computational domain.}
  \label{fig:kdvinfin}
\end{figure}

\begin{figure}
  \centering
  \subfigure[KdV]{%
  \includegraphics[width=0.79\textwidth]{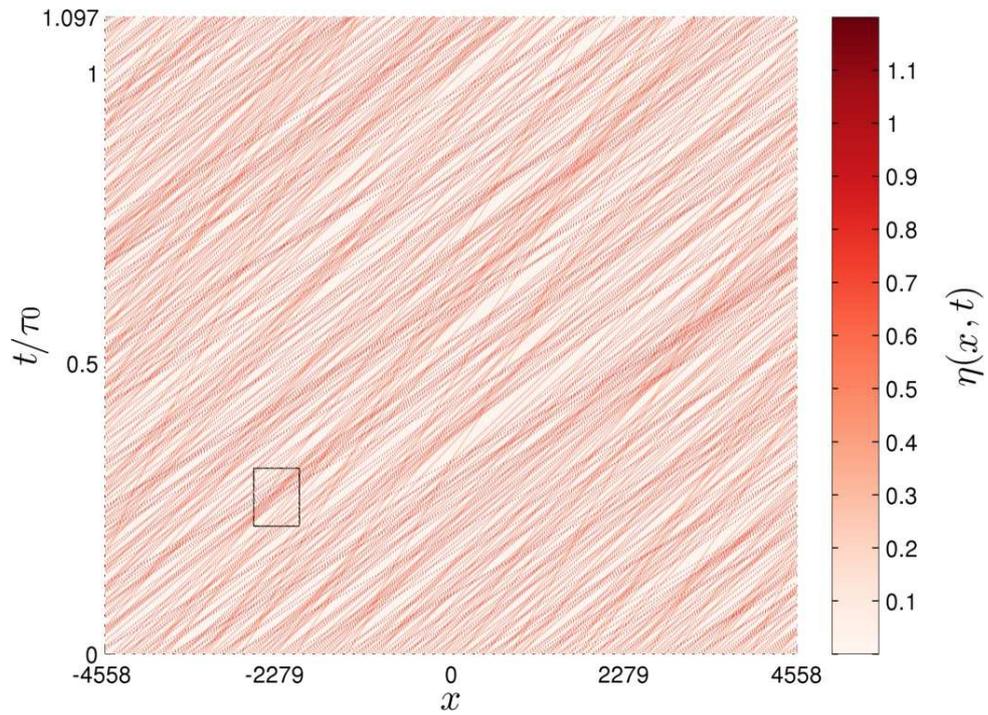}}
  \subfigure[KdV--BBM]{%
  \includegraphics[width=0.79\textwidth]{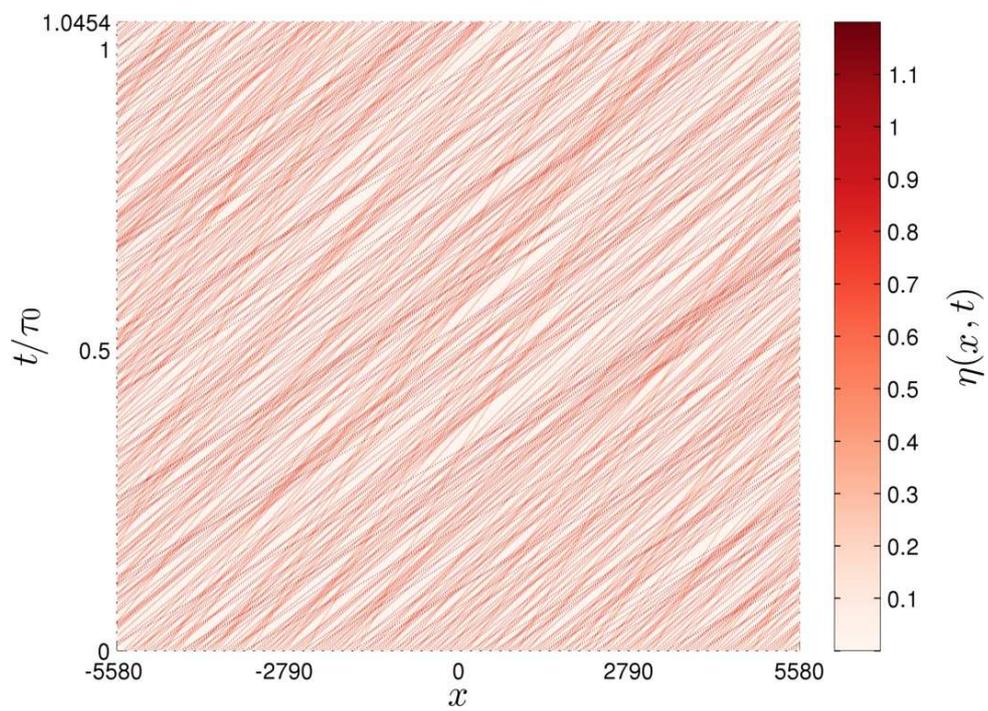}}
  \caption{\small\em Space-time plot of a random solitonic gas under the KdV (a) and KdV--BBM (b) dynamics. The time arrow is directed upwards (the initial state corresponds to the bottom line). The rectangular box in the upper image (a) shows the area zoomed in the next Figure~\ref{fig:spacetime_zoom}.}
  \label{fig:spacetime}
\end{figure}

\begin{figure}
  \centering
  \includegraphics[width=0.69\textwidth]{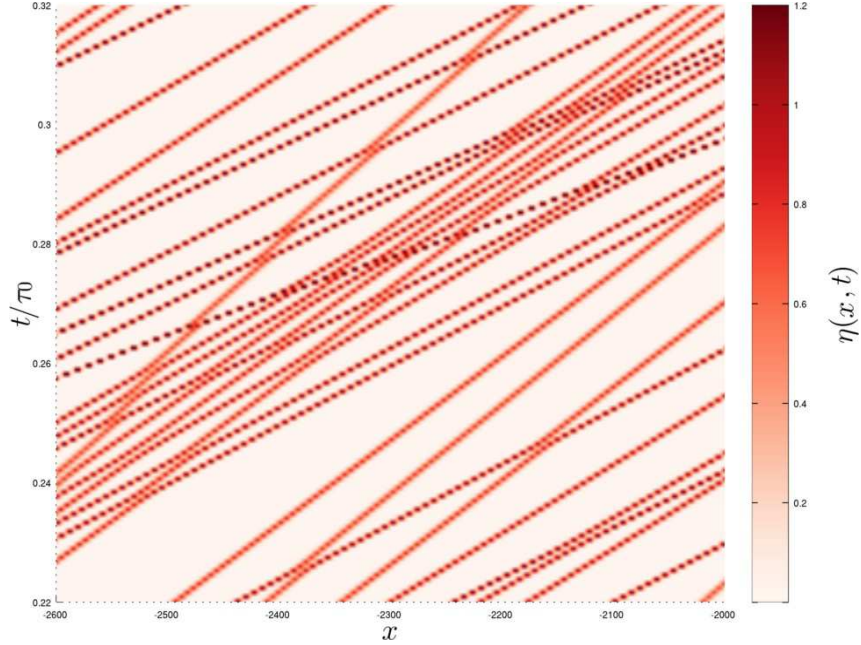}
  \caption{\small\em Zoom on space-time domain $(x,t/\tau_0)\in[-2600, -2000]\times[0.22, 0.32]$ simulated under the KdV equation dynamics. See Figure~\ref{fig:spacetime} for the whole picture.}
  \label{fig:spacetime_zoom}
\end{figure}

The initial condition for the KdV equation is composed of a finite number $N_s$ of solitons with random ampmlitudes $a_i\sim \N(a, \sigma)$, $i=1,\ldots, N_s$ separated by quasi-uniform distance $\Delta x_s$ which is randomized to improve the ergodocity of the initial state:
\begin{equation*}
  \Delta x_s\, :=\, \langle\Delta x_s\rangle + \N(0, \sigma_2),
\end{equation*}
where $\langle\Delta x_s\rangle$ denotes the mean value reported in Table~\ref{tab:params} and $\N(\mu, \sigma)$ is the normal distribution with the mean $\mu$ and the variance $\sigma$. The solitonic gas initial state generated in this way is depicted on Figure~\ref{fig:kdvinfin}(a). We use the same parameters for the initial solitonic gas state for the simulation with the KdV--BBM equation except for the domain size and the spacing between two solitons (see Table~\ref{tab:params}). They are larger in the KdV--BBM case since the soliton width increases with the parameter $\delta \geq 0$ (see Figure~\ref{fig:shapes}). Consequently, we can say that two initial conditions are \emph{approximatively} isomorphic up to the horizontal coordinate stretching transformation. The simulation times $T_1$ (KdV) and $T_2$ (Kdv--BBM) are chosen so that an \emph{average} soliton has enough time to go around the whole computational domain. The final states of the simulations are shown on Figures~\ref{fig:kdvinfin}(b,c). One can notice how the initially quasi-uniform distribution of solitons is mixed by forming instantaneous soliton clusters as long as some void spaces due to the mass conservation property. The complete space-time dynamics simulated using the KdV and KdV--BBM equations is depicted on Figures~\ref{fig:spacetime}(a,b). Individual lines correspond to solitons trajectories. The convergence of these lines corresponds to solitons collisions. It might appear on Figure~\ref{fig:spacetime} that collisions involve multiple solitons, however it is not the case. A zoom on a portion of the space-time domain is shown on Figure~\ref{fig:spacetime_zoom}. One can see that interactions are only binary in agreement with \cite{Zakharov1971}. It is interesting to estimate the total number of collisions in our simulation. The models under consideration are unidirectional. Thus, we can experience only overtaking collisions. Since the simulation time is chosen so that almost every soliton has enough time to travel across the whole computational domain, the number of collisions scales with $\O(m^+ m^-)$, where $m^\pm$ is the number of solitons which travel with the speed above (below) the average. By construction of the initial condition we have $m^\pm = \O(\half N_s)$. Consequently, the total number of collisions scales with $\O(\fourth N_s^2)$.

\begin{figure}
  \centering
  \subfigure[KdV]{%
  \includegraphics[width=0.59\textwidth]{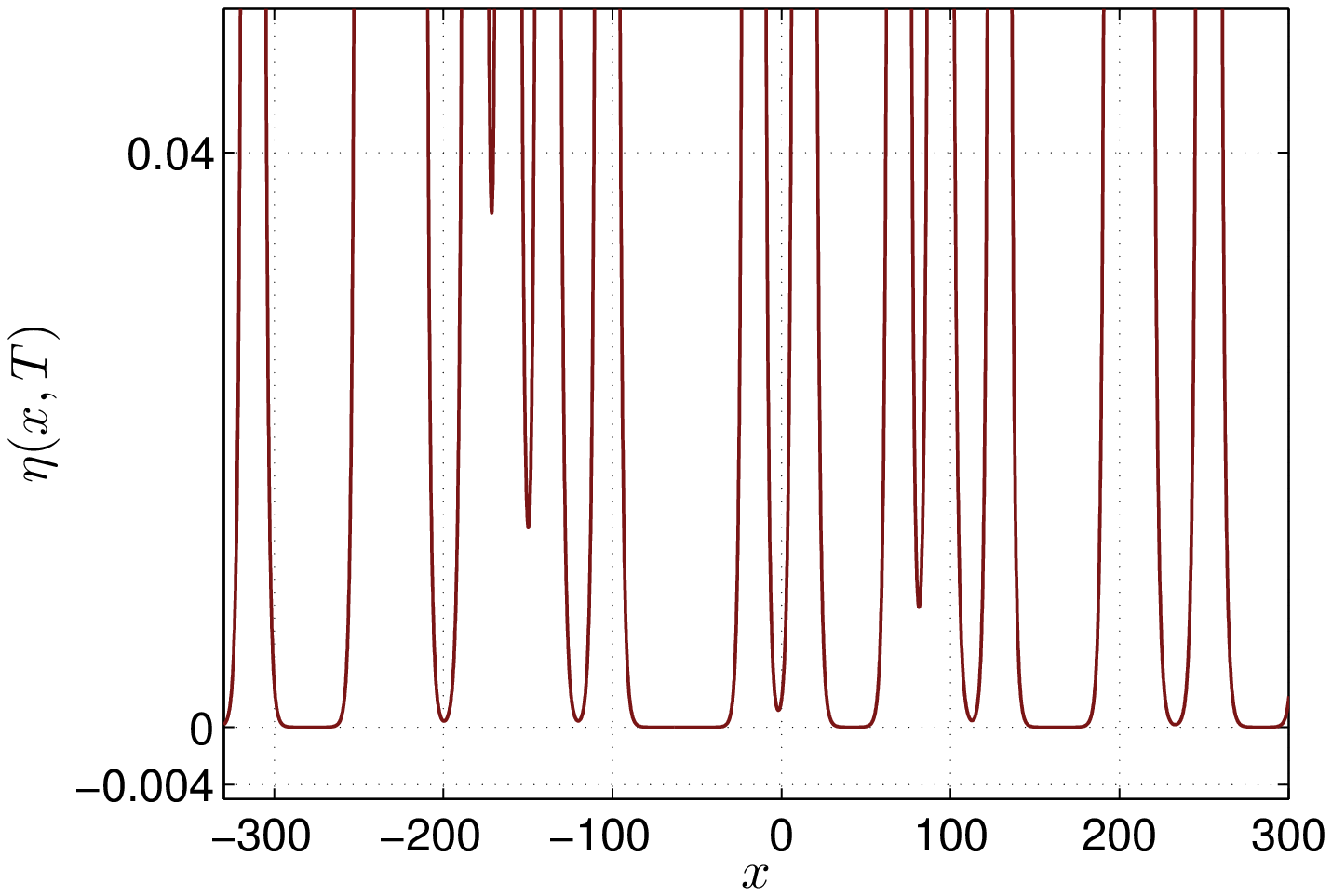}}
  \subfigure[KdV--BBM]{%
  \includegraphics[width=0.59\textwidth]{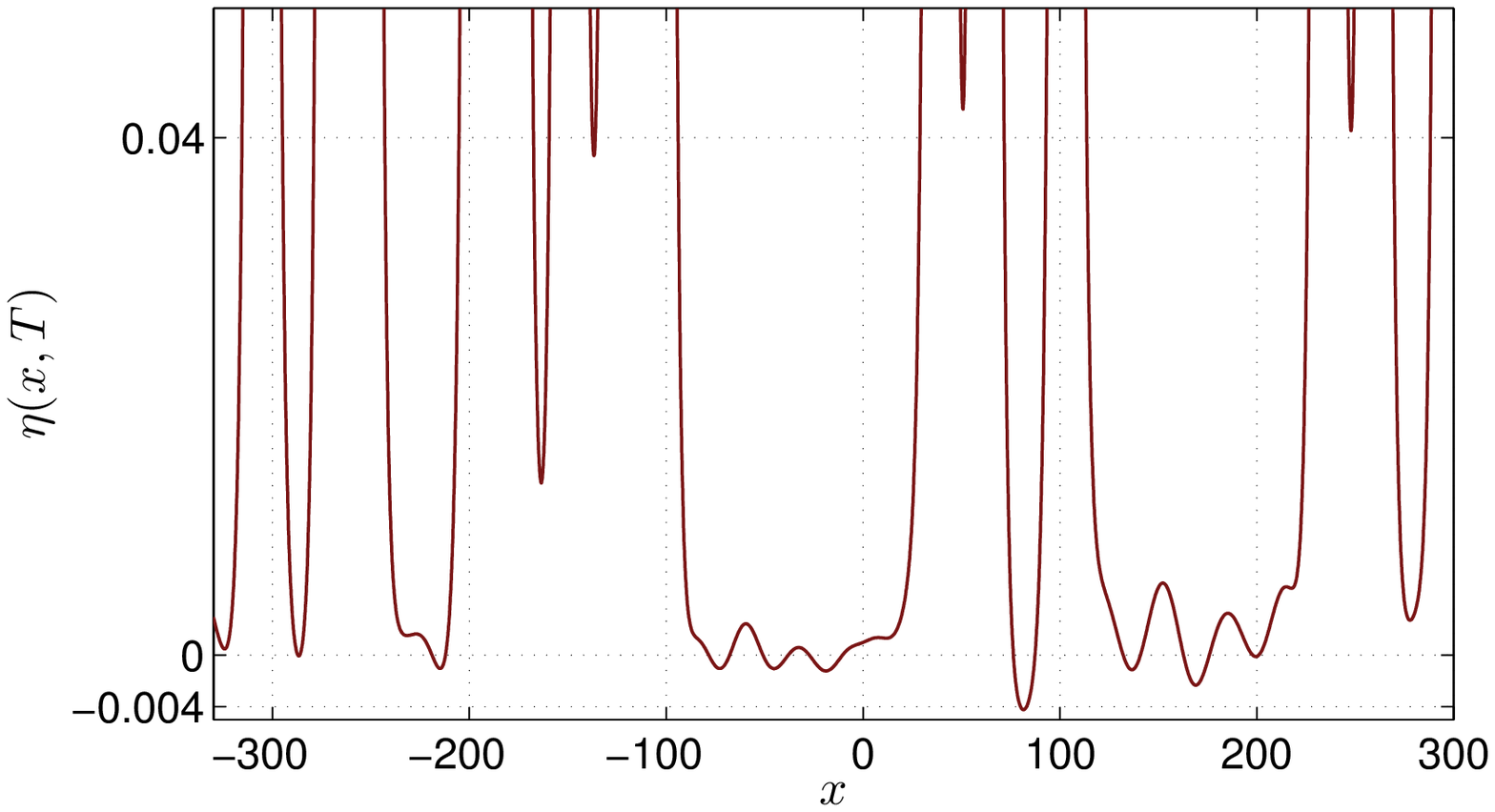}}
  \caption{\small\em Zoom on a portion of the computational domain at the final time of the KdV and KdV--BBM simulations. (b) Tiny oscillations between the solitary waves correspond to the radiation created by inelastic collisions in the KdV--BBM equation.}
  \label{fig:zoom}
\end{figure}

Since the initial conditions are approximatively self-similar, after an appropriate rescaling of the spatial and time variables, the space-time dynamics is similar in both simulations (see Figures~\ref{fig:spacetime}(a,b)). The difference between two simulations can be noticed if one makes a zoom on solitons background in order to see small radiating oscillations due to the inelasticity of collisions in the KdV--BBM case. This zoom on a portion of the computational domain is shown on Figure~\ref{fig:zoom}. In contrast, Figure~\ref{fig:zoom}(a) shows the absence of phonon modes in the KdV simulation.

\begin{figure}
  \centering
  \includegraphics[width=0.69\textwidth]{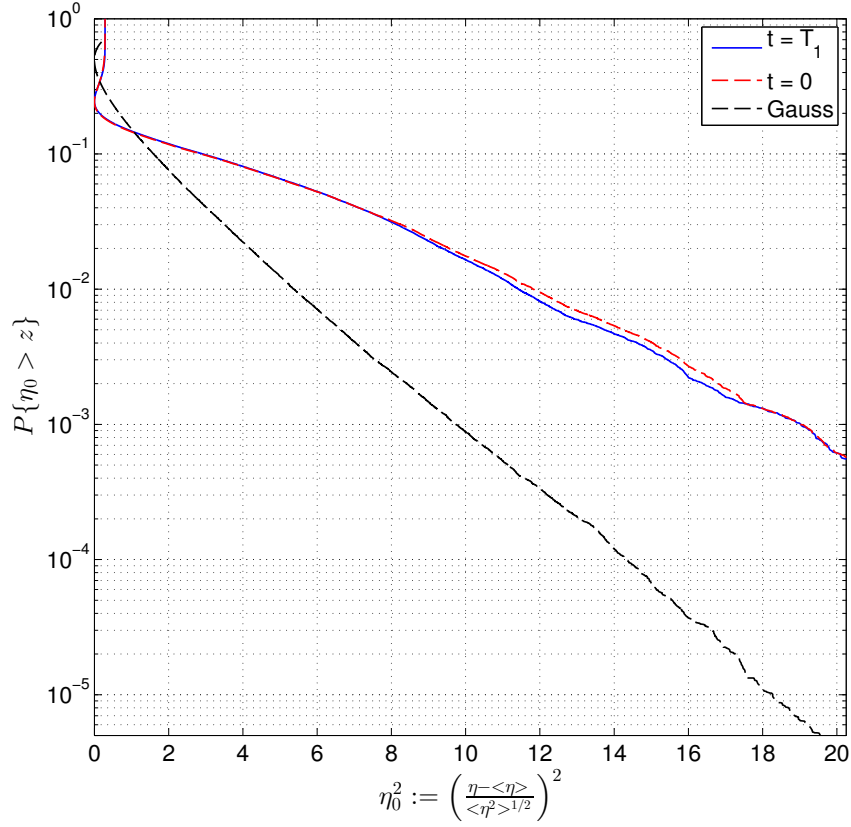}
  \caption{\small\em Probability distributions of the squared normalized free surface elevation.}
  \label{fig:proba}
\end{figure}

\begin{figure}
  \centering
  \subfigure[Kurtosis --- KdV]{%
  \includegraphics[width=0.47\textwidth]{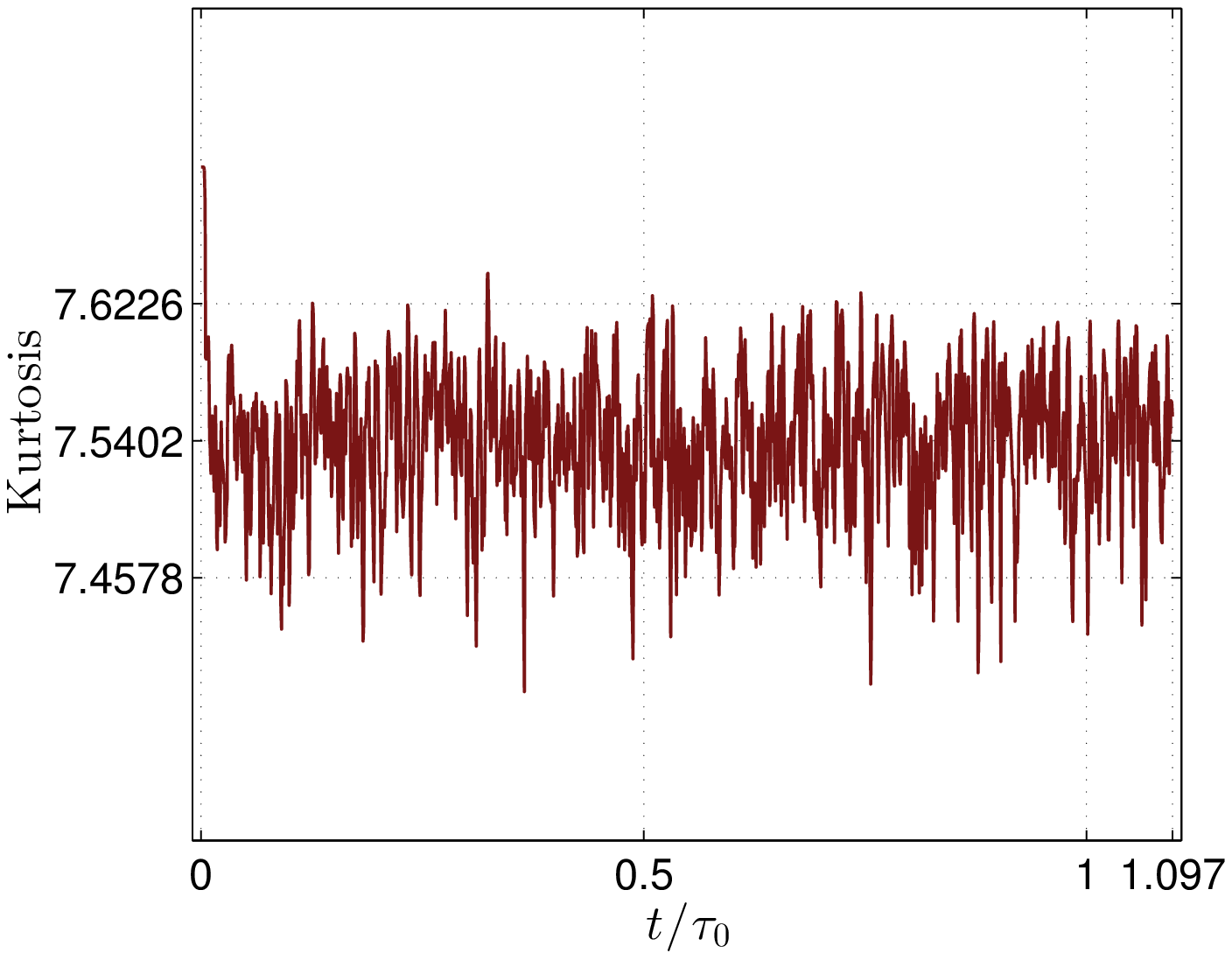}}
  \subfigure[Kurtosis --- KdV--BBM]{%
  \includegraphics[width=0.47\textwidth]{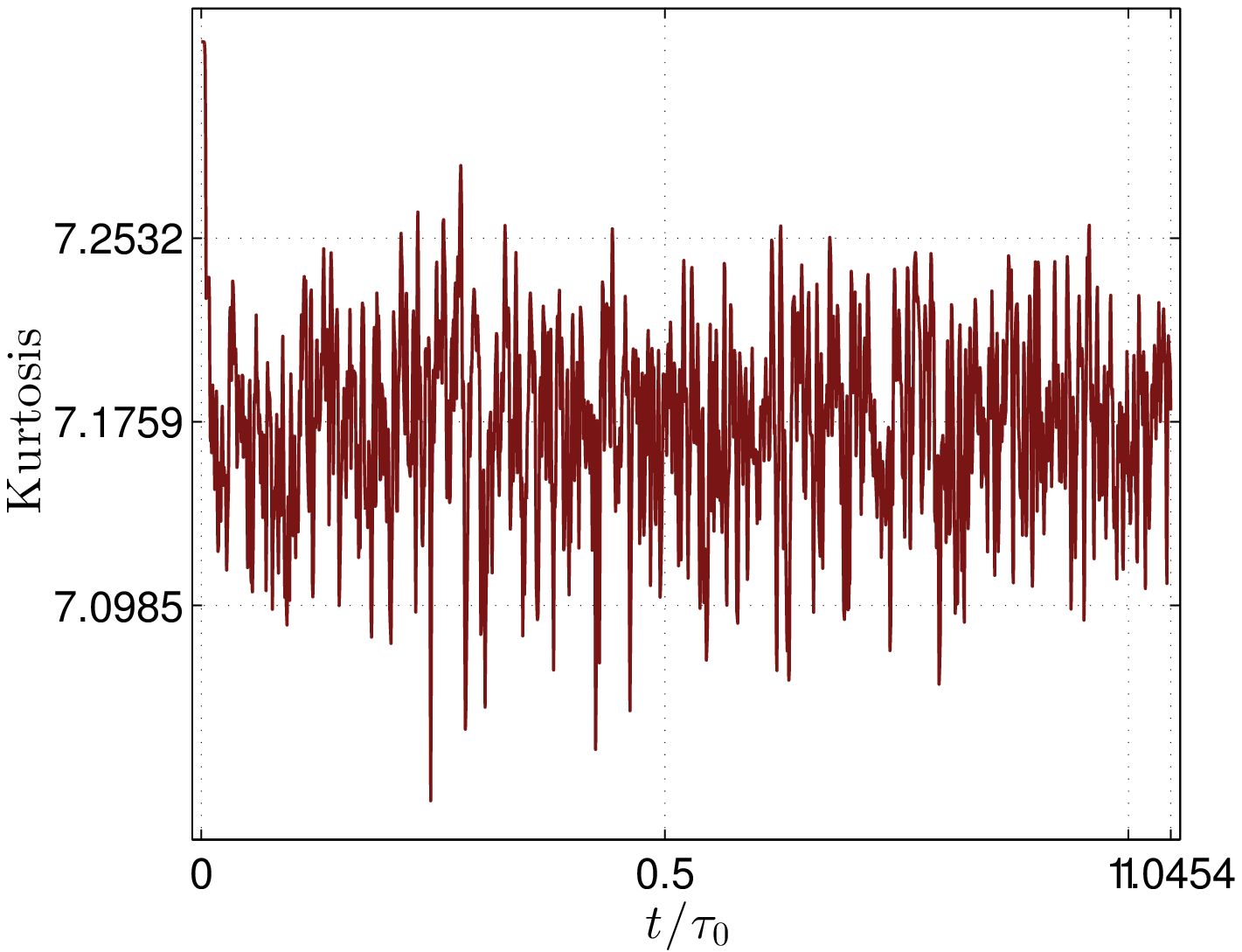}}
  \subfigure[Skewness --- KdV]{%
  \includegraphics[width=0.47\textwidth]{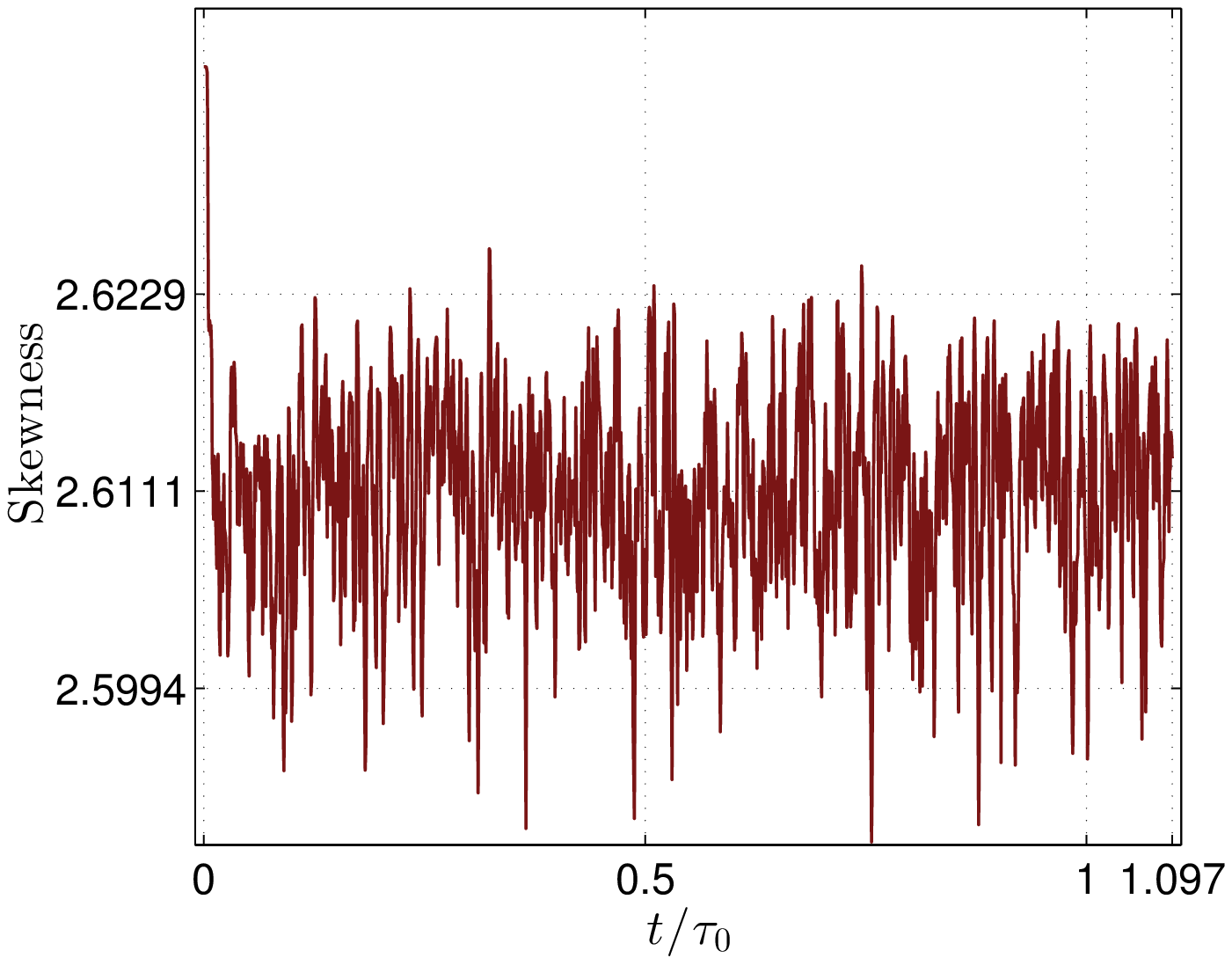}}
  \subfigure[Skewness --- KdV--BBM]{%
  \includegraphics[width=0.47\textwidth]{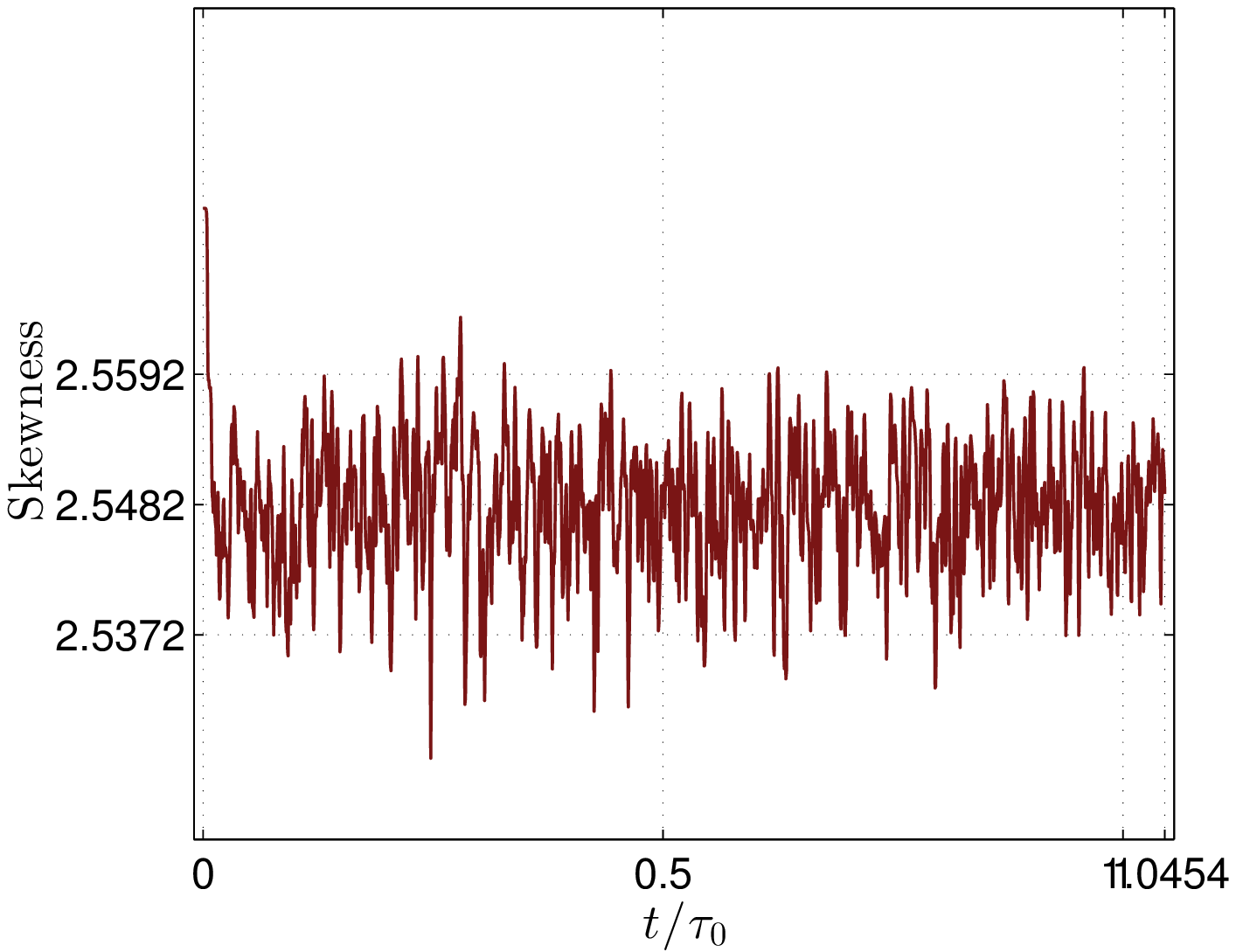}}
  \caption{\small\em Evolution of the kurtosis and skewness in numerical simulations of the KdV and KdV--BBM equations.}
  \label{fig:skewkurt}
\end{figure}

It is custom to use the statistical methods to describe random wave fields \cite{Boccotti2000, Goda2010}. The probability distribution of the normalized free surface elevation $\eta_0(x,t) := \bigl(\eta(x,t) - \langle\eta\rangle\bigr) / \langle\eta^2\rangle^{1/2}$ at times $t = 0$ and $t = T_1$ is shown on Figure~\ref{fig:proba} (under the KdV dynamics). One can see that this distribution is quasi-stationary which is a direct consequence of the fact that solitons preserve perfectly their shape during the interactions. We note that the KdV--BBM numerical result shows the same invariance property since the inelasticity is too weak to modify significantly the probability distribution. Moreover, for comparison we plot also on the same Figure the standard normal (Gaussian) distribution. One can see that numerical results show much heavier tails than the standard distribution depicted on the same plot.

The probability distribution can be characterized by several parameters. Perhaps two most important characteristics are listed hereinbelow:
\begin{itemize}
  \item Kurtosis $\k := \displaystyle{\frac{\mu_4}{\mu_2^2}}$, which measures the \textit{heaviness} of the spectrum tail.
  \item Skewness $\S := \displaystyle{\frac{\mu_3}{\mu_2^{3/2}}}$, which measures the asymmetry of the spectrum with respect to the mean.
\end{itemize}
These quantities are defined in terms of the normalized free surface elevation moments $\mu_n := \langle \eta_0^n \rangle$. We note that $\k = 3$ and $\S = 0$ for the normal (Gaussian) distribution $\N(0, 1)$. The evolution of these quantities is shown on Figure~\ref{fig:skewkurt}. The kurtosis $\kappa$ is shown on top panels (a,b) and the skewness $\S$ on the bottom (c,d). The KdV simulation results are represented on the left images (a,d) and the KdV--BBM on the right (b,d). One can see that the qualitative behaviour of these quantities is similar in integrable and nonintegrable cases. After a rapid initial transient period both quantities $\k$ and $\S$ enter in a quasi-stationary regime which consists of fast small amplitude ($\pm 1.6$\%) oscillations around the mean value. Initial values of the moments are higher, but then it drops down quickly due to soliton interactions. We note also that corresponding values of $\k$ and $\S$ are slightly lower in the KdV--BBM case due to the differences in solitary wave shapes.

\subsection{Estimation of statistical moments}\label{sec:stat}

In order to estimate efficiently kurtosis and skewness for various values of parameters $\St$ and $\delta$ without performing a series of direct numerical simulations we will adopt an approximate analytical method of statistical moments estimation employed also in \cite{Shurgalina2012}.

Let us introduce the average density of a solitonic gas:
\begin{equation*}
  \rho := \frac{N_s}{2\ell},
\end{equation*}
where $N_S$ is the number of solitons and $\ell$ is the half-length of the computational domain (see also Table~\ref{tab:params}). We will assume that the solitonic gas is rarefied, i.e. $\rho \ll 1$. Under this assumption we can represent approximatively the instantaneous free surface elevation $\eta(x,t)$ as a linear superposition of distinct solitary waves (the interacting part is neglected):
\begin{equation*}
  \eta(x,t) \approx \sum_{i=1}^{N_s} \eta_i(x,t) = \sum_{i=1}^{N_s}
  a_i\,\sech^2\bigl(\half\kappa_i\xi\bigr), \qquad
  \xi := x - c_{si}t - x_i,
\end{equation*}
where $\{a_i\}$, $\{c_{si}\}$ and $\{x_i\}$ are respectively the amplitudes, speeds and phase shifts of individual solitary waves. By assuming that the supports of solitons do not overlap, we can estimate the statistical moments of any order. For the sake of simplicity let us consider the first order $\mu_1$, i.e. the mean:
\begin{equation*}
  \mu_1 = \langle \eta(x,t)\rangle = \frac{1}{2\ell}\sum_{i=1}^{N_s} \int_{-\ell/2}^{\ell/2}\eta_i(\xi)\,\ud\xi \approx \frac{1}{2\ell}\sum_{i=1}^{N_s} a_i\int_{-\infty}^{+\infty}\sech^2\bigl(\half\kappa_i\xi\bigr)\,\ud\xi = 4\rho\langle\frac{a_i}{\kappa_i}\rangle,
\end{equation*}
where we took the limit $\ell\to +\infty$ in order to compute analytically the integrals. Note also that the solitary wave parameter $\kappa_i$ is a function of the amplitude $a_i$ according to the relations given in \eqref{eq:sw}.

Higher order moments $\mu_n = \langle\eta^n\rangle$, $n > 1$ can be computed in a similar way. In this study we will need the moments up to the fourth order:
\begin{equation*}
  \mu_2 = \frac{8}{3}\rho\langle\frac{a_i^2}{\kappa_i}\rangle, \qquad
  \mu_3 = \frac{32}{15}\rho\langle\frac{a_i^3}{\kappa_i}\rangle, \qquad
  \mu_4 = \frac{64}{35}\rho\langle\frac{a_i^4}{\kappa_i}\rangle.
\end{equation*}
Using these moments we can estimate the skewness $\S$ and kurtosis $\k$ in the rarefied gas limit $\rho\to 0$:
\begin{equation}\label{eq:skew}
  \S = \frac{\langle(\eta - \mu_1)^3\rangle}{\langle(\eta - \mu_1)^2\rangle^{3/2}} = \frac{\mu_3}{\mu_2^{3/2}} + \O(\rho^{1/2}) \approx \frac{2\sqrt{3}}{5\sqrt{2}}\rho^{-1/2}\frac{\langle a_i^3/\kappa_i\rangle}{\langle a_i^2/\kappa_i\rangle^{3/2}},
\end{equation}
\begin{equation}\label{eq:kurt}
  \k = \frac{\langle(\eta - \mu_1)^4\rangle}{\langle(\eta - \mu_1)^2\rangle^2} = \frac{\mu_4}{\mu_2^2} + \O(1) \approx \frac{9}{35}\rho^{-1}\frac{\langle a_i^4/\kappa_i\rangle}{\langle a_i^2/\kappa_i\rangle^2}.
\end{equation}

In order to validate these asymptotic expressions we compare their predictions for the solitonic gas described in Section~\ref{sec:num}. The skewness $\S$ and kurtosis $\k$ are computed from numerical simulations and the time average is then taken. The results of the comparison are provided in Table~\ref{tab:comp}. One can see that the simple analytical model presented in this Section is clearly able to describe rarefied solitonic gases. Moreover, one can infer from \eqref{eq:skew}, \eqref{eq:kurt} the behaviour of the skewness and kurtosis as the density of the gas decreases $\rho\to 0$. However, as $\rho\to 0$, the jump from the initial value of $\S(0)$ or $\k(0)$ to the stationary one will diminish, since the interactions between solitons become sparser.

\begin{table}
  \centering
  \begingroup\setlength{\fboxsep}{0pt}
  \colorbox{lightgray}{
  \begin{tabular}{r|c|c}
  \hline\hline
           & Skewness, $\S$ & Kurtosis, $\k$ \\
  \hline
  KdV ($\delta = 0$)     & $\mathbf{2.6111} / 2.6365$ & $\mathbf{7.5402} / 7.7047$ \\
  \hline
  KdV--BBM ($\delta = 2$) & $\mathbf{2.5482} / 2.5732$ & $\mathbf{7.1759} / 7.3359$ \\
  \hline\hline
  \end{tabular}}\endgroup
  \bigskip
  \caption{\small\em Comparison of the numerical (\textbf{left}) to the approximate analytical (right) estimations for the skewness \eqref{eq:skew} and kurtosis \eqref{eq:kurt} on the data analysed in the previous section.}
  \label{tab:comp}
\end{table}

The asymptotic formulas \eqref{eq:skew}, \eqref{eq:kurt} can be used to estimate the skewness $\S$ and kurtosis $\k$ for various values of model parameters $\delta$ and $\St$ in initial stages of the solitonic gas evolution. However, these formulas contain the statistical averages. The proposed Monte--Carlo approach is briefly summarized here. Namely, we generate $M$ random independent realizations of a solitonic gas, each sample consisting of $N_s$ solitons. The numerical values of parameters $M$ and $N_s$ used in simulations are given in Table~\ref{tab:params}. Then we use the assumption of the rarefied gas along with the knowledge of the analytical soliton shape in order to estimate some statistical quantities of the gas (see equations \eqref{eq:skew}, \eqref{eq:kurt}, for example). A big number of Monte--Carlo realisations allows to annihilate local fluctuations and to obtain robust estimations.

\begin{figure}
  \centering
  \includegraphics[width=0.69\textwidth]{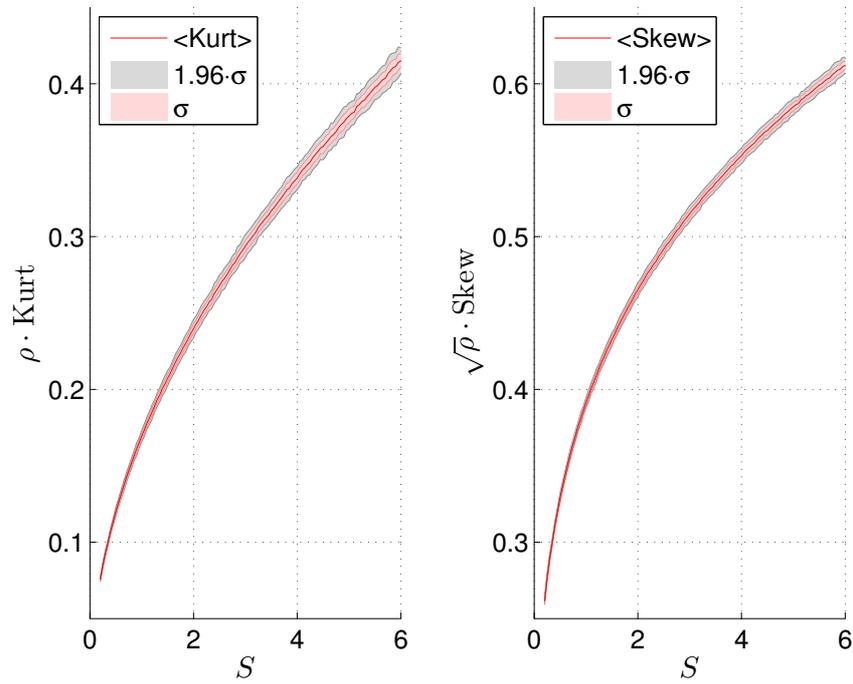}
  \caption{\small\em Initial kurtosis (left) and skewness (right) dependence on the Stokes--Ursell number $\St$.}
  \label{fig:mcstokes}
\end{figure}

\begin{figure}
  \centering
  \includegraphics[width=0.69\textwidth]{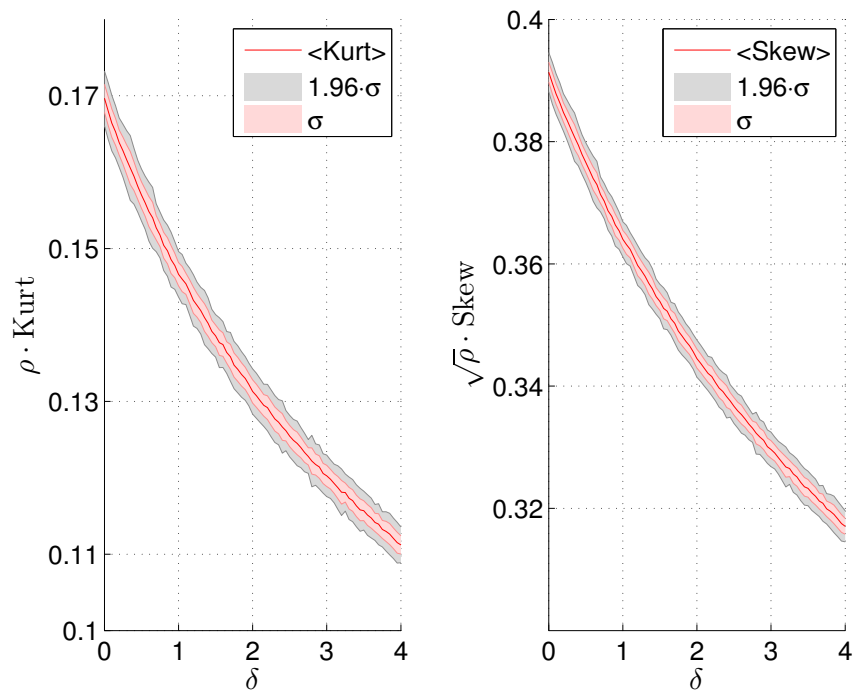}
  \caption{\small\em Initial kurtosis (left) and skewness (right) dependence on the BBM-term coefficient $\delta$.}
  \label{fig:mcdelta}
\end{figure}

\begin{figure}
  \centering
  \includegraphics[width=0.69\textwidth]{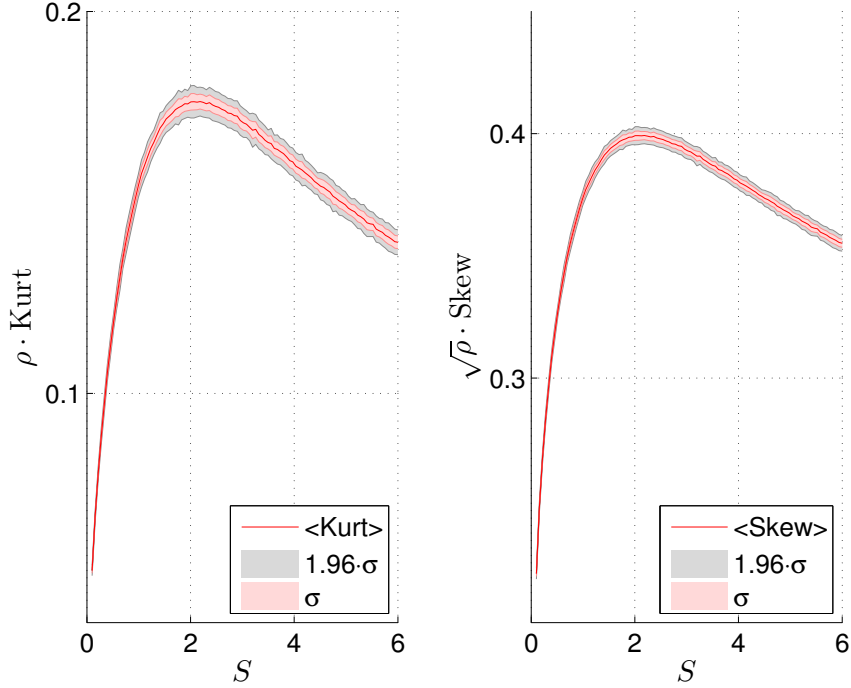}
  \caption{\small\em Initial kurtosis (left) and skewness (right) behaviour when one increases simultaneously $\delta$ and $\St$.}
  \label{fig:mcstdelta}
\end{figure}

The numerical results are presented on Figures~\ref{fig:mcstokes} -- \ref{fig:mcstdelta} where the shadowed areas show the statistical error due to Monte--Carlo simulations ($\pm\sigma$, $\pm 1.96\sigma$, where $\sigma$ is the estimated variance). From these results one can clearly see that for the fixed density $\rho$ an increase in the Stokes--Ursell number $\St$ leads to an increase in $\S$ and $\k$ (see Figure~\ref{fig:mcstokes}). The BBM coefficient $\delta$ has the antagonistic effect as it is illustrated on Figure~\ref{fig:mcdelta}. When both parameters $\St$ and $\delta$ are increased simultaneously\footnote{It means that we introduce an auxiliary homotopy parameter $\lambda\in [0,1]$ which parametrizes the simultaneous change of parameters $\St$ and $\delta$ in the following way:
$$
  \St = \St_{\min}(1-\lambda) + \St_{\max}\lambda, \qquad
  \delta = \delta_{\min}(1-\lambda) + \delta_{\max}\lambda.
$$
In the computations presented below we took the values: $\St_{\min} = 0.1$, $\St_{\max} = 6.0$, $\delta_{\min} = 0.0$ and $\delta_{\max} = 4.0$.}, the dependence of statistical quantities is not monotonic anymore (see Figure~\ref{fig:mcstdelta}).

\section{Conclusions and perspectives}\label{sec:concl}

In this study we presented several numerical experiments on the solitonic gas turbulence in the framework of an integrable KdV and a nonintegrable regularized KdV--BBM equation. The numerical results reported above generalize previous investigations \cite{Pelinovsky2006, Dutykh2013a} where only a limited number (a few dozens) of solitons were used to represent a solitonic gas. Consequently, we reduce the statistical error according to the law of large numbers.

First of all, we showed that the probability distribution for the solitonic gas remains quasi-invariant during the system evolution for both KdV and KdV--BBM cases. The special attention was payed to the statistical characteristics such as kurtosis and skewness which measure the `\emph{heaviness}' of tails and the asymmetry of the free surface elevation distribution. In particular, using the asymptotic methods and Monte--Carlo simulations we showed that both skewness and kurtosis increase with the Stokes--Ursell number $\St$ and decrease when the BBM term coefficient $\delta$. When both parameters $\St$ and $\delta$ are increased gradually and simultaneously, these effects are in competition: first we observe the increase of these statistical characteristics, but then, this tendency is inversed and they decrease after reaching their respective maximal values (see Figure~\ref{fig:mcstdelta}). We would like to underline that the proposed Monte--Carlo methodology is much less computationally expensive than direct numerical simulations. Despite the small number of Monte--Carlo runs ($M = 100$) the estimated statistical error is sufficiently small for the purposes of this study. On the other hand, this approach is restricted, strictly speaking, to the situations where the solitons are well separated, since it does not take into account superpositions of solitons.

The present study opens a number of perspectives for future investigations. More general nonlinearities could be included into the model along with some weak dissipative and forcing effects. This could allow us to observe Kolmogorov spectra of a solitonic gas \cite{Zakharov1992}. The nonintegrable effects need some time to be accumulated. Consequently, even longer simulation times are needed. The interaction of a solitonic gas with a random radiation field has to be studied as well.

\subsection*{Acknowledgments}
\addcontentsline{toc}{section}{Acknowledgments}

D.~\textsc{Dutykh} acknowledges the support from ERC under the research project ERC-2011-AdG 290562-MULTIWAVE. E.~\textsc{Pelinovsky} would like to thank for the support the VolkswagenStiftung, RFBR grants (14-05-00092). The authors would like to thank Gennady \textsc{El} for stimulating discussions on the topics of integrability and solitons theory.

\begin{center}
  \line(1,0){150}
\end{center}

\addcontentsline{toc}{section}{References}
\bibliographystyle{abbrv}
\bibliography{biblio}

\end{document}